\documentclass[letter,superscriptaddress]{revtex4}
\usepackage{epsfig}
\usepackage{graphicx}

\begin{document}
\title{\bf
The shortcomings of semi-local and hybrid functionals: what we can learn
from surface science studies.
}

\author{A. Stroppa}
\email{alessandro.stroppa@univie.ac.at}

\author{G. Kresse}
\affiliation{Faculty of Physics, University of Vienna, and Center for Computational Materials Science,
 Sensengasse 8/12, A-1090 Vienna, Austria}

\begin{abstract} 
 A study of the adsorption of CO on late $4d$ and $5d$ transition metal (111) surfaces
 (Ru, Rh, Pd, Ag, Os, Ir, and Pt) considering atop and hollow site adsorption is presented. 
 The applied functionals include the gradient corrected PBE and BLYP functional, 
 and the corresponding hybrid Hartree-Fock density functionals HSE and B3LYP.
 We find that PBE based hybrid functionals (specifically HSE) yield, 
 with the exception of Pt, the correct site order
 on all considered metals, but they also considerably overestimate the adsorption energies
 compared to experiment.  On the other hand, the semi-local BLYP functional 
 and the corresponding hybrid functional B3LYP yield very satisfactory adsorption energies
 and the correct adsorption site for all surfaces. We are thus faced with
 a Procrustean problem: the B3LYP and BLYP
 functionals seem to be  the overall best choice for describing adsorption on metal surfaces, 
 but they simultaneously fail to account well for the properties of the metal, 
 vastly overestimating the equilibrium volume and underestimating the atomization energies.
  
 Setting out from these observations, general conclusions are drawn on the relative merits
 and drawbacks of various semi-local and hybrid functionals. The discussion includes a revised
 version of the PBE functional specifically optimized for bulk properties and
 surface energies (PBEsol), a revised version of the PBE functional specifically
 optimized to predict accurate adsorption energies (rPBE), 
 as well as the aforementioned BLYP functional. 
 We conclude that no semi-local functional is capable
 to describe all aspects properly, and including non-local exchange also
  only improves some, but worsens other properties.
\end{abstract}


\maketitle

\section{Introduction}
\label{sec:intro}

The adsorption of carbon monoxide has intrigued researchers in surface science
for the last decades. This has obvious reasons. Carbon monoxide is a simple
diatomic molecule, but despite its simplicity it shows a rich phase diagram
on metal surfaces.\cite{COphasediagram,Santen4}
The CO dissociation reaction involves a very simple reaction mechanism, but nevertheless the
reaction barriers
can be greatly varied by alloying or roughening the surface.\cite{Santen4}
It is fair to say that, together with hydrogen adsorption and dissociation,
CO has become a classical benchmark for experimental and theoretical
surface science techniques. This interest is also strongly driven by the importance of CO for
many technologically relevant reactions.\cite{Gabor}

In this light, it is unsatisfactory  that state of the art density functional
calculations fail to describe several aspects of the adsorption of
CO on metal surfaces accurately. This concerns both the predicted
adsorption site as well as the absolute magnitude of the adsorption
energy. For the Cu, Rh and Pt (111) surface, the  generalized gradient
approximation (GGA) in the PBE version predicts that CO adsorbs at a  high coordination
site (typically the hollow site), whereas experiments unequivocally
show that atop adsorption is preferred. For Ag and Au a degenerate
site preference is predicted, which does not agree with
experiments either.\cite{CO_Pt,Hafner} Equally compelling is that
adsorption energies are significantly overestimated using the PBE
functional.\cite{CO_Pt,Hafner} These discrepancies have first been
noticed in a now classical paper by Feibelmann {\em et al.} and the
name \emph{CO adsorption puzzle} has been coined.\cite{CO_Pt} Since
then there has been emerging evidence that the present local and semi-local 
functionals are not capable to describe the subtle balance between donation of
charge to the substrate and back donation to the molecule correctly
for many adsorption problems.  
However, it is still an open question whether this is a universal shortcoming
of such functionals, or whether the functionals can be amended to improve
the description of surface related properties without drastically worsening
other important properties.\cite{Alaei}

The common model to describe CO adsorption is the  Blyholder model,\cite{Bly}
which invokes interactions of the two CO frontier orbitals, the
 $5\sigma$ HOMO (highest occupied molecular orbital) and the $2\pi^*$
 LUMO (lowest unoccupied molecular orbital),
with the metal states.
Due to the interaction with the metal states bonding  $5\sigma$-metal
orbitals and anti-bonding $5\sigma$-metal orbitals develop, and the latter
are partly shifted above the Fermi-level of the metal, causing a net
bonding interaction (donation). Likewise, bonding  $2\pi^*$-metal hybrid states become
populated (back-donation).
Simple symmetry reasons tell that the highly directional $5\sigma$-metal interaction is
particularly strong for atop adsorption, whereas the $2\pi^*$ interaction is dominating
 for hollow site adsorption.\cite{Santen1,Santen2,Santen3,Santen4,Kresse1,Gil,way1}

Here we apply a set of now well established semi-local as well as hybrid functionals to the
CO adsorption problem. Our focus is on the systematic variation across the periodic
table, and on how hybrid functionals compare to the more traditional semi-local
functionals.
We show that none of the available functionals is
capable to  yield an equally satisfactory description of the {\em metal and the
CO adsorption}.

\section{Theoretical and computational method}
\label{sec:model}

The first principles density functional theory calculations and the
hybrid functional calculations
have been
performed using the VASP code, PAW\cite{PAW} potentials in the implementation
of Kresse and Joubert\cite{PAWKresse} and a cutoff energy of 400 eV.
The surfaces are modelled by a periodic
four layer metal slab with $c(2\times 4)$ symmetry, and
the CO molecule is adsorbed on one side of the slab.
We focus only  on the atop and hollow fcc adsorption sites, since
  the difference in calculated adsorption energies between the hcp and fcc
 hollow adsorption site is generally small, on the order
of 0.05 eV; see, e.g.,
Refs.~\onlinecite{fccvshcp1,fccvshcp2,Stroppa07,Stroppa08_1}. The Brillouin zone
integration is performed on  symmetry  reduced $\Gamma$-centered
6$\times$6$\times$1 grids (i.e. roughly $12\times 12\times 1$ in the
primitive cell). Since we want to
concentrate on trends, 
all metals are considered in the fcc structure,
although Os and Ru crystallize in the hcp lattice structure. 

The PBE\cite{PBE} and BLYP\cite{BLYP1,BLYP2} functionals are used
for the GGA calculations. For the hybrid functional calculations,
the HSE functional\cite{HSE03,ErrorReport,HSE06} has been applied. 
We use a variant of the  HSE06 functional that observes the homogeneous electron
gas limit and all important sum rules. Contrary to the
conventional HSE06 functional, the screening parameter is set
to $\omega=0.300~$\AA$^{-1}$ in both the semi-local as well as non-local part
of the exchange functional, whereas the recommended choice is
$\omega=0.207~$\AA$^{-1}$. As has been  shown previously, the specific
choice of the screening parameter $\omega$ has very little influence
on the total energies, but slightly affects band gaps.\cite{HSE06,Paier2,Paier2Erratum}
Throughout this paper we will use the acronym HSE for this functional. As we have shown elsewhere, 
for Cu, Rh and Pt the HSE  functional yields the
same site order and almost the same energetics as the more popular hybrid
PBEh  (sometimes also termed PBE0 or PBE1PBE) functional\cite{PBE01,PBE02} 
(see Ref.~\onlinecite{Stroppa07} for details).

Furthermore the widely adopted
B3LYP\cite{B3LYP1,BLYP1} functional is applied in the present work.
The difference to the previous functional is predominantly in the
correlation energy which is approximated using the semi-local
Lee-Yang-Parr correlation. Systematic tests for selected extended
systems are described elsewhere.\cite{Paier2,Paier2Erratum,JCP04,JCP05,PRB06} 
Implementational details can be found in Ref. \onlinecite{Paier2}.
For the PBE and HSE functionals, the
bulk metal lattice constants were optimized and used in the periodic
slab calculations. In order to save compute time, the BLYP and B3LYP
calculations were performed at the HSE geometries, which also agree best with experiment.

\section{Results}
\subsection{Lattice constants}
\label{sec:lattice}

\begin{table}
 \caption{
Lattice constants $a$(\AA) of late $4d$ and $5d$ transition metals
in the fcc structure using different
functionals. Experimental lattice constants of Ru and Os have been
calculated from the experimental hcp  volume assuming an fcc
structure. Experimental values are taken from
Ref.~\onlinecite{Webelements}. Numbers in round brackets are the
relative
 error (in percent) with respect to the experimental lattice constants.
The experimental lattice constants  have not been extrapolated to 0 K, and the zero-point
quantum fluctuations have not been included in the calculations.
Inclusion of these contributions will change the values by approximately 
$\sim$0.1~\% \cite{lattconsts}.}
\label{tab:lattice}
\begin{ruledtabular}
\begin{tabular}{l|ccccc}
              &   EXP &        HSE      &        PBE      &   B3LYP        & BLYP         \\
\hline
Ru            & 3.79  &  3.76 ($-$0.8)  & 3.80 (+0.3)   &  3.82 (+0.8)   & 3.87 (+2.1)   \\
Rh            & 3.80  &  3.78 ($-$0.5)  & 3.82 (+0.5)   &  3.85 (+1.3)   & 3.90 (+2.6)   \\
Pd            & 3.89  &  3.93 (+1.0)    & 3.94 (+1.3)   &  3.99 (+2.6)   & 4.04 (+3.8)   \\
Ag            & 4.09  &  4.14 (+1.2)    & 4.15 (+1.5)   &  4.23 (+3.4)   & 4.27 (+4.4)   \\
Os            & 3.82  &  3.82 (+0.0)    & 3.85 (+0.8)   &  3.87 (+1.3)   & 3.92 (+2.6)   \\
Ir            & 3.84  &  3.84 (+0.0)    & 3.88 (+1.0)   &  3.90 (+1.5)   & 3.95 (+2.9)   \\
Pt            & 3.92  &  3.93 (+0.2)    & 3.97 (+1.3)   &  4.02 (+2.5)   & 4.06 (+3.6)   \\
\end{tabular}
\end{ruledtabular}
\end{table}

The lattice constants of the considered metals are summarized in table~\ref{tab:lattice}.
We do not report the bulk moduli, since a one to one relation between them
and the volumes exists (see e.g. Ref.~\onlinecite{Paier2,PaierB3LYP} for the PBE, HSE, and B3LYP case):
if the lattice constant is too small, the bulk modulus is overestimated, and
\emph{vice versa}.
 Hence the quality of the functional can be assessed
by the predicted lattice constants. For Cu
(Ref.~\onlinecite{Stroppa07,Stroppa08_1}), Rh, Ag, and Pt we can
compare the PBE and BLYP lattice constants calculated using three
different codes, e.g. VASP (this work),
 Quantum-ESPRESSO and Wien2k (Ref.~\onlinecite{Alaei}): in all cases, the  agreement
  is very good typically within a few 0.1 \%, except for Rh where the discrepancy between our PBE value and that in
 Ref.~\onlinecite{Alaei} calculated using Quantum-ESPRESSO is 1.4 \%.
We relate this to the pseudopotential approximation applied in Ref. ~\onlinecite{Alaei}.

 Table~\ref{tab:lattice}  shows
that the overall description is best on the HSE level, good on the
PBE level, at best, modest using the B3LYP functional and
quite bad using BLYP.
The HSE functional always yields smaller lattice constants than
the PBE functional, and the contraction is largest for  half-filled $d$ bands
and smallest when the $d$ band is entirely filled (Ag). Compared to
experiment, this yields to a sizeable
underestimation of the lattice constants for Ru, a sizeable
overestimation for Ag and Pd, and fairly accurate values
for the remaining elements.
The B3LYP functional gives  larger lattice constants with
substantial errors  for Ag, Pd and Pt (3.4~\%, 2.6~\% and 2.5~\%).
Finally, the  BLYP functional yields the largest lattice constants with 
 errors up to 4~\% for Pd, Ag and Pt.
 Paier {\em et al.} have already shown in Ref.~\onlinecite{PaierB3LYP} that this
is mainly related to significant errors in the LYP correlation
functional for metallic systems. 

In summary, ad-mixing non-local exchange decreases the
lattice constants, whereas the LYP correlation gives larger lattice
constants than the PBE correlation.
\subsection{Metal $d$ band parameters}

\label{sec:parameters}

\begin{table}
 \caption{$d$ band parameters for the $4d$ and $5d$ VIII B transition metals.
  $\epsilon_d$ is the center of the $d$ band, $n_{d}$ is the $d$ band filling, $\Phi$ is the
 work function of the clean (111) surface, calculated using PBE. Numbers in round brackets
 refer to HSE calculations. $\Phi_{\rm exp}$ is the experimental work
 function.\cite{workfunctions}  $\Phi_{\rm exp}$ of Ru and Os refers
 to the hcp(0001) surface.}
\label{tab:parameter}
\begin{ruledtabular}
\begin{tabular}{l|ccc}
                     &   Ru           &Rh             &    Pd          \\
\hline
$\epsilon_{d}$ (eV)  &$-$2.59 ($-$3.17)& $-$2.37 ($-$2.83) &$-$1.84 ($-$2.09)    \\
$n_{d}$              & 6.50           &  7.57         & 8.71            \\
$\Phi$(eV)           & 5.1 (4.9)      &  5.4 (5.0)    & 5.4 (5.3)       \\
$\Phi_{\rm exp}$(eV)     & 4.7            &  5.0          &    5.6          \\
\hline
                     &  Os            & Ir            &   Pt            \\
$\epsilon_{d}$ (eV)  &$-$3.01 ($-$3.60)  &$-$2.93 ($-$3.39) & $-$2.45 ($-$2.88)   \\
$n_{d}$              &  6.20          & 7.24          & 8.33            \\
$\Phi$(eV)           & 5.4 (5.4)      &   5.5 (5.3)   &5.7 (5.6)        \\
$\Phi_{\rm exp}$(eV)     &  4.8           &  5.7          &    5.7          \\
\end{tabular}
\end{ruledtabular}
\end{table}

In table~\ref{tab:parameter} we show the calculated metal $d$ band
parameters using PBE and HSE: the center of the $d$ band,
$\epsilon_{d}$ and the metal work function, $\Phi$. The experimental
work function is also reported. The $d$ band center can not be
defined unambigously
 within  a plane-wave approach. As we are mainly interested in trends
 (\emph{i.e.} PBE vs. HSE),  we choose to calculate
 the centre of gravity of the
occupied $d$ band (integration up to the Fermi level),
which is a lower bound for the true $d$ band
center (see Ref.~\onlinecite{Hafner} for details).
In table~\ref{tab:parameter}, we have also introduced a \emph{fractional}  $d$ band
filling ($n_{d}$), calculated as the $d$ charge contained
in a  sphere centered at the metal atom.
The sphere radius has been fixed in order to
contain  {\em exactly} (within numerical error estimated to be 0.02 $e$)
 the  number of valence $s$ and $d$ electrons in the atomic
configuration.\cite{LTRUNC}
Interestingly, it turns out that the volume of the sphere is almost
equal to the volume per atom in the fcc lattice (Wigner Seitz sphere), 
providing a physically sound basis for our definition of $n_{d}$.

 We considered the \emph{bulk}
phase\cite{bulkphase, Filling} at equilibrium geometry using the PBE
functional for calculating $n_{d}$. We will show below that this
definition is helpful for discussing trends between and among the
$4d$ and $5d$ metals.
From table~\ref{tab:parameter}, we see that
$\epsilon_{d}$ calculated using HSE
($\epsilon_{d}^{\textrm{HSE}}$)
is always found at larger binding energies than
 $\epsilon_{d}$ calculated using PBE ($\epsilon_{d}^{\textrm{PBE}}$). This is related to the tendency
 of   HSE to give a larger $d$ bandwidth and a slightly weaker self-interaction within the $d$ shell 
than PBE.~\cite{Stroppa07} The
common picture of reactivity of metal surfaces suggests that
the lower in energy   $\epsilon_{d}$ is, the smaller the 
adsorption energies should be. We will show below that this is in contradiction
with our results for HSE. Although
$\epsilon_{d}^{\textrm{HSE}}<\epsilon_{d}^{\textrm{PBE}}$, the adsorption energies
\emph{generally} increase for HSE (see figure~\ref{fig:Eads01}).
\label{sec:charge}

The HSE work function is generally smaller than the one calculated
using PBE. This can be traced back to the induced surface dipole
changes due to the inclusion of the Fock exchange. Indeed, hybrid functionals
 cause  a complex redistribution of the  charge density in the surface layer: 
 an analysis of the difference
 between the PBE and HSE charge density (not shown here) shows that HSE functional 
 depletes charge from the $d_{z^{2}}$ orbital in favor of  $d_{xz}$  and
$d_{yz}$ orbitals, thus decreasing the spill-out of  electrons
from the surface layer into the vacuum. This charge redistribution should also
affect the initial steric repulsion, thus explaining why, for the
top site, the HSE adsorption energies are larger than the PBE ones
(see figures~\ref{fig:Eads01},~\ref{fig:Eads02} and ~\ref{fig:Eads2}).
The net effect of this complex charge
redistribution is to reduce slightly the surface
dipole and the work function.\cite{workchanges}
 The overall agreement with the experimental work function is good
for both functionals.

\subsection{The gradient corrected PBE functional}
\label{sec:gradient}

\begin{figure}
\includegraphics[width=8cm,angle=0,clip=true]{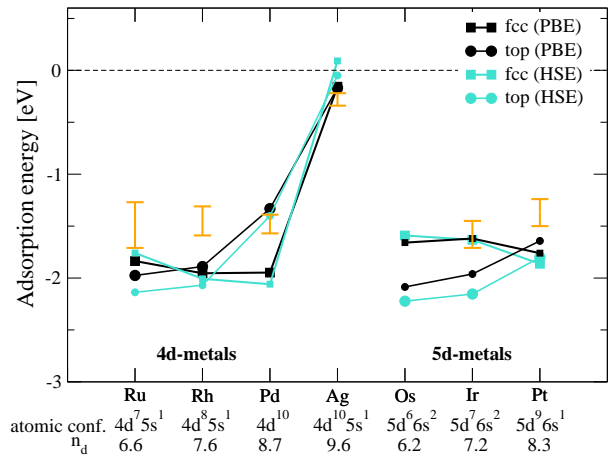}
\caption{Adsorption energies for the atop and fcc hollow
sites  on late $4d$ and $5d$ metals for PBE and HSE.
Experimental data with error bars are taken from
Ref.~\onlinecite{corr1}. 
The experimental adsorption site at low coverage 
is  hollow for Pd and top for the other cases.} \label{fig:Eads01}
\end{figure}

\begin{figure}
\includegraphics[width=8cm,angle=0,clip=true]{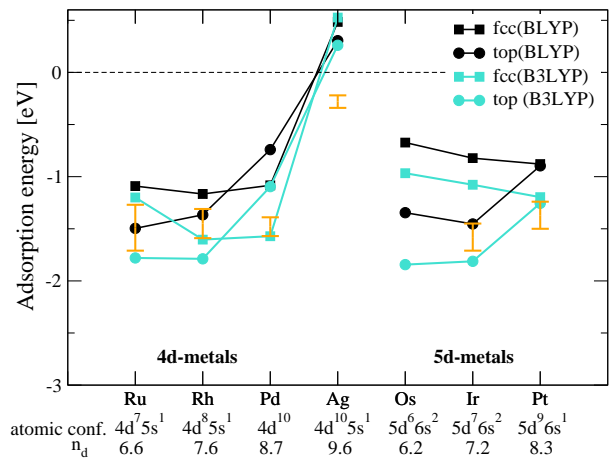}
\caption{Adsorption energies for the atop and fcc hollow
sites  on late $4d$ and $5d$ metals for BLYP and B3LYP.
Experimental data with error bars are taken from
Ref.~\onlinecite{corr1}.} \label{fig:Eads02}
\end{figure}
The central results of the present work are summarized in figures~\ref{fig:Eads01} and \ref{fig:Eads02},
where we show the adsorption energies for the atop and fcc site, using  PBE, HSE, BLYP and B3LYP.
For B3LYP and BLYP, they are evaluated using the HSE equilibrium geometries. 
Experimental values are also shown with error bars.\cite{corr1}
We will first concentrate on the PBE results, which have been published in a similar
form by Gajdos \emph{et al.}\cite{Hafner}
For atop adsorption, the adsorption energy progressively decreases towards
the noble metals. The qualitative and to some extend quantitative behavior
 of the atop curve can be
reproduced using the Hammer-N\o rskov $d$ band model,\cite{HammerNorskov,HammerNorskov1}
 in which  the central parameter
is the position of the $d$ band with respect to the Fermi-level.
 It moves to lower energies as the $d$ band
becomes  filled. A word of caution is in order here. Using our definition, the $d$ band 
moves to higher energy as the $d$ band filling increases (table~\ref{tab:parameter}).
 This is not in contradiction to 
previous findings,\cite{HammerNorskov1} since the different behaviour is related to a different 
definition of the $d$ band center. In Ref.~\onlinecite{HammerNorskov1} the \emph{full} $d$ band is considered,
while here we focus only on the occupied portion. The same trends as found in table~\ref{tab:parameter}
have been reported  in Ref.~\onlinecite{Hafner}. As we are mainly interested in trends, our following 
discussions are not affected by the particular definition. 

\begin{figure}
\includegraphics[width=10cm,angle=-90,clip=true]{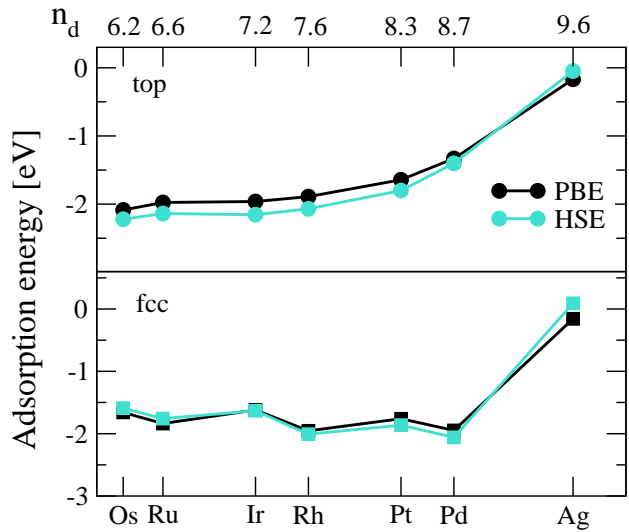}
\caption{Adsorption energies for the atop and hollow site adsorption on
late $4d$ and $5d$ metals shown versus the $d$ band filling.
}
\label{fig:Eads2}
\end{figure}

We note, however, that models based on the weak
interaction limit, which forms the  basis for the derivation of the
Hammer-N\o rskov approach, possess only limited
value in the case of the strong interaction present for CO adsorbed
on a transition metal surface, in particular, at the hollow site.\cite{Santen4} In fact,  the $d$ band model describes
well the overall trends in the periodic table, but it does not
account for the site dependent geometric effects (\emph{i.e.}  top
vs. hollow).
A convenient way to present the results for the $4d$ and $5d$
transition metals is to show all values in a single graph  and to
choose as the abscissa the occupation of the $d$ band
(figure~\ref{fig:Eads2}) according to the previous definition. Interestingly
for the top site, all adsorption energies fall on a single curve,
suggesting that for top site adsorption the filling of the $d$ band
and not the position of the $d$ band is the most sensible 
parameter (descriptor) to characterize the interaction strength.\cite{Santen4}
Figure~\ref{fig:Eads2} clearly shows a
 decreasing preference for atop binding in the sequence Os$<$Ru$\sim$Ir$<$Rh$<$Pt$<$Pd.
These findings agree with experimental evidence within
experimental errors (see Ref.~\onlinecite{corr1} for further
details) and other theoretical calculations.\cite{Koper} Furthermore,
 for many properties, for instance catalytic
properties, Pt behaves similar to a fictitious element between Rh
and Pd, as the $d$ band filling is roughly in-between the latter two
elements. According to conventional wisdom, the $5\sigma$ interaction
dominates for atop coordination. Hence, the decrease 
of the adsorption energy with increasing $n_{d}$ filling can be 
explained in terms of  occupation of antibonding $5\sigma-d$
metal states that destabilize the molecular surface bond.

Quantum chemical calculations
have shown  that  the back donation mechanism is
  the most dominant for the hollow site (2$\pi^{*}-d$ interaction).\cite{Santenback1,Santenback2}
We find, in agreement with previous studies,\cite{Hafner} that
the hollow site adsorption energy generally {\em increases} towards the late
transition metals, with a final sharp decrease for the noble metals.
This decrease for the noble metals is related to a fully occupied
$d$ band, located several eV below the Fermi-level. Therefore, 
for noble metals it becomes difficult to shift antibonding $5\sigma-d$
states above the Fermi-level, and additionally the interaction
between the deep and contracted $d$ orbitals and the $2\pi^*$
frontier orbitals becomes small. It turns out that sensible 
descriptors are the work function and the $d$ band filling. In fact,
previous studies have shown that the back-donation interaction is
very sensitive to the relative position of the $2\pi^{*}$ orbitals
and the $d$ band, which also depends on the metal work
function  $\Phi$.\cite{Koper, Koper1} Clearly, a larger $\Phi$ value at
fixed or almost constant $d$ filling ({\em i.e.} Os vs Ru; Ir vs Rh;
Pt vs Pd) implies a downward shift of the Fermi  energy, decreasing
the back-donation into the $2\pi^{*}-d$ bonding orbitals,
thereby decreasing the fcc adsorption energy. As a result, the CO adsorption at
the fcc site is less stable on Os, Ir and Pt than on  Ru, Rh and Pd.
When increasing the $d$ filling, that is moving from left to right
along the periodic table, the back-donation is also strengthened: since the
bonding $2\pi^{*}-d$ orbitals become increasingly occupied, 
the molecular surface bond is stabilized. Counteracting this trend  is the change
in the work function, which increases towards the late transition metals.
Finally, steric repulsion, which is always stronger at
the hollow site than at the atop site, also
plays a role. It decreases when moving to the right of the
periodic row.\cite{Concepts}  Although we found it impossible
to disentangle the aforementioned factors, the calculations show
that the combined effect of these trends is that hollow site adsorption becomes stronger
towards the late transition metals.
Last, but not least, we want to note
that the different qualitative trends for the atop and fcc
adsorption energies as a function of the $d$ filling reflect the different 
dominant interaction term  at the atop and fcc site, that is donation and back-donation,
respectively.

We now return to figures~\ref{fig:Eads01} and ~\ref{fig:Eads02}. It is clearly born out that
the top site is preferred for early transition metals, but moving
along the transition metal series the hollow site becomes
progressively more stable. For the PBE functional, a change in site
preference is observed between Ru and Rh for the $4d$ metals, and Ir
and Pt for the $5d$ metals, earlier than observed in experiment.
The noble metals behave somewhat differently. As
already argued above, this is related to a very weak $5\sigma \to d$
donation and weak $d \to 2\pi^*$ back-donation. Here bonding is
dominated  by the $s$ and $p$ orbitals, but again the PBE functional
fails to predict the right site order.
A 0.1 eV shift of the fcc and top curves 
against  each other would give the correct site
preference  for all metals, e.g. top for Rh and Pt, but maintaining hollow site adsorption
for Pd 

\subsection{The hybrid HSE functional}
\label{sec:hybrid}

The hybrid HSE  functional shifts the $2\pi^*$ orbitals towards the
vacuum level and the  $5\sigma$ orbital towards stronger binding
energies.\cite{Stroppa07} In combination with a downshift of the $d$ band center,
caused by a reduced self-interaction within the $d$ shell due to the
inclusion of Fock-exchange, one might naively believe that the
interaction between the $2\pi^*$ orbitals and the metal $d$ states
should weaken for hybrid functionals. Although this conjecture is
confirmed for the noble metals, for transition metals the weaker
interaction is counteracted by the much increased $d$ band width of
the metal and a concomitant decrease of the work-function from PBE to the HSE
functional (see table~\ref{tab:parameter}).

It is clear from both, figures~\ref{fig:Eads01},~\ref{fig:Eads02} and ~\ref{fig:Eads2}, that
 the adsorption energy {\em increases} for most adsorption sites using HSE:
(i) the increase is pronounced and roughly constant for the top site
on the transition metals; (ii) for the hollow site a slight
destabilization is visible for Os and Ru, and a slight stabilization
is visible for Pd and Pt.  It is remarkable how smooth the changes
along the transition metal series are, supporting the high precision
of the present calculations and excluding numerical artifacts. 
As a result of these shifts, the crossing points between the hollow and top site
move towards the right, for
both $4d$ and $5d$ metals, but not sufficiently so to favor atop adsorption on Pt(111).
For the other critical cases, i.e. Rh, Ag (and Cu, not included here but discussed
in Ref. \onlinecite{Stroppa07}),
HSE is capable to recover the correct site order, but only just. The main reason why HSE fails
to predict CO adsorption above the top Pt site is that the hollow site also
becomes more stable for the late transition metals, in particular Pd and Pt, whereas
the increase in stability  is small at the hollow site for earlier transition metals.
This increased stability may be related to the tendency of HSE
to increase surface energies (see below) and the tendency to \emph{localize} the charge
at the metal atoms, causing charge \emph{depletion} at the hollow sites.
\begin{figure}
\includegraphics[width=7cm,angle=0,clip=true]{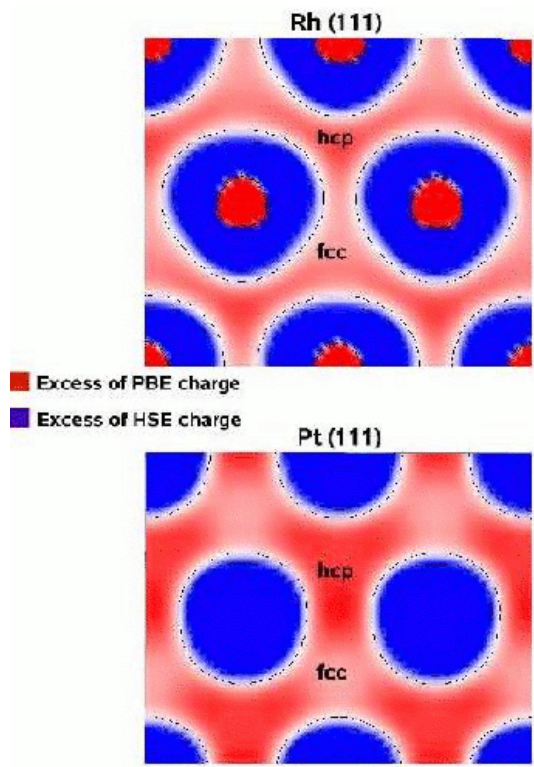}
\caption{Difference between the PBE and HSE
charge density for Rh and Pt \emph{clean} surfaces plotted in a (111) plane  $\sim$ 0.7~\AA~above the
surface plane (\emph{i.e.} between the C atom of CO adsorbed at the hollow site and the surface metal atoms).
Blue (red) color corresponds to regions where the PBE charge density is larger (smaller) than the HSE one.
Three isolines are drawn to guide the eyes. ``fcc" and ``hcp" indicates the  location
of the fcc and hcp hollow site, respectively.}
\label{fig:hollow}
\end{figure}
In figure~\ref{fig:hollow} we show  the difference between the PBE and
HSE charge density, plotted for a (111) plane $\sim$ 0.7 ~\AA~
above the   metal atoms of the clean surface, that is, roughly in between the CO
molecule and the metal surface layer. The figure indeed indicates charge 
depletion above the hollow site (red regions), suggesting a decrease of the steric repulsion using HSE.
Furthermore, figure~\ref{fig:hollow} suggests that, going from PBE to
HSE, the hcp site should increase more in stability than the fcc site.
This is confirmed by our previous calculations for Rh\cite{Stroppa07}
that gave a stabilization of $\Delta E^{\rm PBE-HSE}_{\rm fcc}=$~0.007 eV for
the fcc site, and 
$\Delta E^{\rm PBE-HSE}_{\rm hcp}=$ 0.027 eV for the hcp site.

The most unsatisfactory result is that the adsorption energies
increase substantially for the transition metals from PBE to HSE, which leads to
fairly strong disagreement with experiment on the HSE level---
even more so than for the PBE functional. 
We note that the same site preference, but  even larger adsorption energies are obtained
using the standard PBEh hybrid functional (also termed PBE0 or PBE1PBE functional).
This functional also includes 25~\% non-local exchange but, contrary to the HSE functional,
does not truncate the long range part of the  non-local exchange.\cite{Stroppa07}
Since the only difference
between the PBE and PBEh functional is that the latter one replaces
25 \% of the local approximation to the exchange by the Hartree-Fock
exchange, the conclusion is that {\em inclusion of 25 \% non-local exchange, 
bare or long-range screened, hardly
improves the description of CO adsorption on metal surfaces.}

This confirms our previous study,\cite{Stroppa07,Stroppa08_1}
but contradicts a recent work, where it is claimed that the CO
adsorption puzzle has been solved for Pt using hybrid
functionals.\cite{Trieste} In order to explain the contradiction,  
we  have calculated the adsorption energies using norm-conserving and PAW potentials
and we present the results in the Appendix. We show
that the conclusions of Ref.~\onlinecite{Trieste} derive from 
 an artifact of the pseudopotential approximation,
and all-electron methods still yield  the wrong site order for CO on Pt(111)
using the PBEh functional.

\subsection{The hybrid B3LYP functional}
\label{sec:B3LYP}
Our results also  somewhat  disagree with other reports, which
have applied the B3LYP functional and found stronger preference for
atop adsorption and weaker adsorption energies on Cu and
Pt.\cite{Gil,Doll,Neef,Scheffler} To resolve this issue,  B3LYP
values are  reported in figure~\ref{fig:Eads02}. In fact, the
B3LYP functional shifts the crossover point even further to the right and
lowers the adsorption energies significantly compared to the HSE
and PBE functional. Since these ``improvements'' can {\em not} be
related to the inclusion of non-local exchange (see above), the
difference must stem from the semi-local part, and in all likelihood
the Lee Yang Parr correlation energy (see e.g.
Ref.~\onlinecite{PaierB3LYP}). In agreement with this conjecture,
the semi-local BLYP functional also predicts the right site order on
all considered metals. In summary, improvements from the PBE to the
B3LYP functional are mostly related to the LYP
correlation functional, and only partly to the inclusion of
non-local exchange, confirming the recent work by Alaei \emph{et
al.}\cite{Alaei}

The seeming ``improvements'' for the BLYP and B3LYP functional over PBE and HSE, however, come
at a considerable price best illustrated in table~\ref{tab:lattice}. The equilibrium
volume of Ag is overestimated by 12 \%, that of Pt and Pd by about 9 \%,
an error that is by no means acceptable.
Somewhat less troublesome but still worthwhile mentioning is that
adsorption on the noble metals becomes slightly endothermic using B3LYP and BLYP
(no geometry optimization was performed for the B3LYP case). We do
not want to overemphasize this point, since presently DFT functionals
generally lack weak van der Waals like dispersion forces, which might contribute
 to the CO bonding on noble metals.

\section{Discussion}
\label{sec:discussion}
\begin{figure}
\includegraphics[width=9cm,angle=0,clip=true]{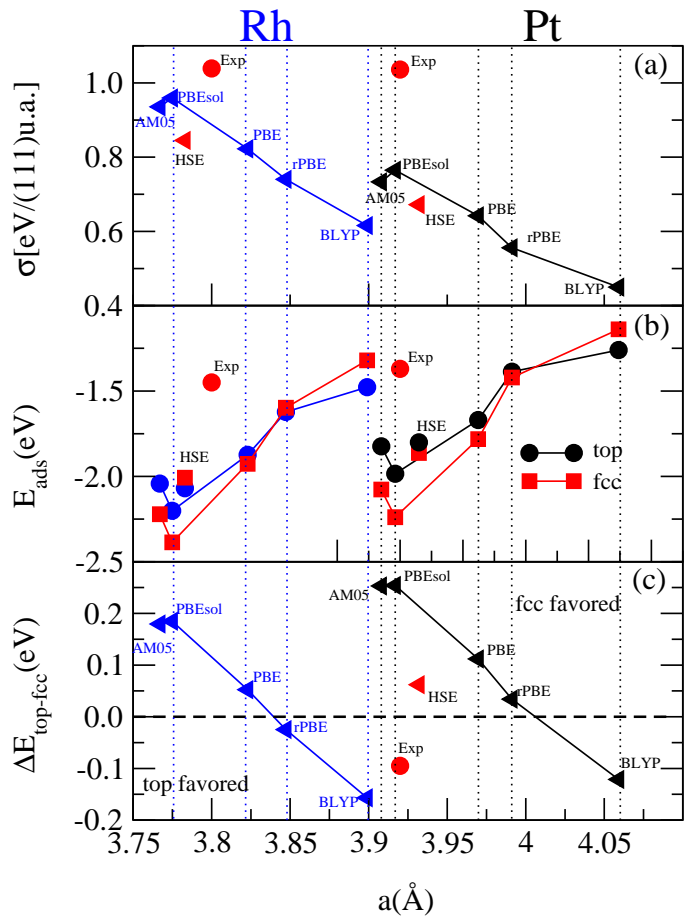}
\caption{Surface energies $\sigma$, adsorption energies E$_{\rm ads}$ for atop and fcc hollow sites,
and energy difference between top and fcc 
sites $\Delta E_{\rm{top-fcc}}$ for CO on Rh and Pt(111) versus theoretical lattice constant $a$ (\AA) 
 calculated
using a set of different functionals. The experimental surface 
and adsorption energy is taken from Ref.~\onlinecite{Vitos}
and Ref.~\onlinecite{corr1} respectively. The experimental $\Delta E_{\rm{top-fcc}}$ 
for CO on Pt(111) has been inferred from Refs.~\onlinecite{De2,De3}. 
Surface calculations were performed at the theoretical lattice constant corresponding to
the applied functional.} \label{fig:trend}
\end{figure}

It has recently been suggested that the quest for improved semi-local functionals
is not yet over, and that it might be possible to design semi-local functionals
that better describe the adsorption on surfaces.\cite{Alaei} 
Unfortunately this is a misconception, more precisely, for
semi-local functionals, it seems impossible
to improve the description of adsorption without worsening other
properties. To discuss this issue we
have extended our study to three additional functionals, the rPBE functional 
by Hammer and N\o rskov,\cite{RPBE} and the recently suggested
PBEsol\cite{PBEsol} and AM05\cite{AM05} functional. The last functional
is based on the subsystem approach and designed to yield accurate surface
energies. The other two functionals are modifications of
the standard PBE functional. These modifications are in the opposite direction:
 the PBEsol functional follows similar ideas as the AM05 functional in order to
improve surface energies, whereas the rPBE functional improves formation energies
and adsorption energies.

Most gradient corrected functionals essentially modify the LDA exchange and
correlation energy to account for areas of large gradients. This is usually
done by increasing the exchange and correlation energies locally
in those areas where the gradient is large. For the ``canonical''
PBE functional, the enhancement factors  have been determined using sum-rules, and the result
is a well balanced functional that describes molecular
and solid state  properties equally well.
There are however failures, in particular small molecules are over-bound, 
whereas solids involving elements beyond the second row are usually under-bound. Resultantly, 
for the metals considered here the 
lattice constants are too large and the surface energies much too small (underbinding).
To highlight this, figure~\ref{fig:trend} plots the surface energy for all considered functionals 
versus the theoretically predicted lattice constants of Rh and  Pt. Similar results are to be expected 
for the other transition metals.

The AM05 functional has been designed to describe jellium and jellium surface energies
accurately. 
The PBEsol functional mimics the AM05 functional by reducing the enhancement factor
in the area of strongly varying gradients, thus approaching
the LDA description.\cite{PBEsol} As a rule of thumb, both 
increase the cost for surfaces, and hence systems involving surface areas become 
less favourable. Atoms are destabilized, 
the formation energies of molecules increase
(molecules possess a smaller surface area per atom than atoms)
and the atomization energies of bulk materials become much larger (no surface area).
Furthermore bulk lattice constants shrink to reduce ``internal" surfaces. 
The AM05 and PBEsol functionals shift the focus towards
a proper description of solids, but in doing so small molecules
are strongly overbound compared to atoms using the PBEsol and AM05 functional.\cite{PBEsol}
In figure~\ref{fig:trend} (a), the AM05 and PBEsol functionals yield 
smaller lattice constants and larger surface
energies than the PBE functional; both are now in quite reasonable agreement
with experiment.

The rPBE functional does exactly the opposite: it increases
the enhancement factor in the area of strongly varying gradients, and it therefore 
moves further away from the LDA description. 
As a rule of thumb,  the rPBE functional decreases the cost for surfaces. Therefore,
molecules are destabilized compared to atoms, bulk materials become less stable than
molecules, and bulk lattice constants increase. The decrease in the surface
energy and increase in the lattice constant is again visible
in figure~\ref{fig:trend} (a). 
The BLYP functional, characterized by an even larger enhancement
factor (see e.g. Ref.~\onlinecite{Alaei}), continues this trend.

We can now understand, why the adsorption energy decreases from 
AM05 and PBEsol, over PBE, rPBE to BLYP [Figure~\ref{fig:trend} (b)]:
when CO adsorbs on the surface, the total surface of the combined
system CO + surface is effectively smaller than that of the
isolated CO and the bare surface.\cite{RPBE} 
In AM05 and PBEsol, surfaces are expensive, and thus adsorption processes, which always lead 
to a reduction  of the effective surface area, yield a larger  energy gain. 
For rPBE and even more so for BLYP, surfaces are
cheap, and less energy is gained during adsorption processes.

The change in the energy difference between the top and the hollow
site [figure~\ref{fig:trend} (c)] can be understood by reasoning along the same line.
Atop adsorption reduces the surface area very little essentially
along a single bond between the CO and the surface, whereas
hollow site adsorption reduces the surface area along three
bonds between the three neighbouring surface atoms and the CO.
If surfaces are ``expensive'' (PBEsol), hollow site adsorption is
preferred, since hollow site adsorption reduces the effective
surface area most. If surfaces are ``cheap'', top
site adsorption is preferred. 

The binding energies, surface energies, lattice constants, 
and adsorption energies are thus {\em all} linearly related, and
we could plot any quantity versus the other and we will always observe
a linear relationship. 
In figure~\ref{fig:trend} we have
chosen the lattice constant as independent variable, but the
surface energy would work equally well (and arguably might be
the better control parameter). Within the family of
gradient corrected functionals, PBE seems to remain the best compromise.
The AM05 and PBEsol functionals describe the lattice constants of the solids best
and also yield improved surface energies, but unfortunately they
overestimate the heats of formation, and important
for the present study, the binding energies of molecules on surfaces.
The BLYP functional arguably yields the best adsorption energies
and the right site order, but for the price of crossly wrong
bulk energies, much too large lattice constants, and  much too small surface 
energies.

The hybrid  functionals constitute a deviation--- albeit only a subtle  one---
from this  general trend. 
The inclusion of exact exchange also clearly increases
the cost for surfaces. Resultantly, HSE leads to a
smaller lattice constant than PBE, and B3LYP to a smaller
lattice constant than BLYP. Even the average
adsorption energies follow roughly the same trend as
for semi-local functionals: in particular the adsorption energy
for the top site
lies  on the same curve as for the semi-local functionals.
The only major deviation is visible for the energy difference
between the top and hollow site. 
This we believe to be related to the shift
of the 2$\pi^*$ LUMO orbital towards the vacuum level, decreasing
the capability of the 2$\pi^*$ orbital to accept charge from
the substrate. As we have argued before this generally
decreases the adsorption energy, with a stronger effect
for the hollow site than the for top site.
On passing, we note that on the basis of surface energy arguments
alone, one would expect hybrid functionals to yield similar
atomization energies for molecules and solids as a
semi-local functional predicting similar lattice constants. This would imply that the atomization energies of
molecules were significantly overestimated and that
of solids slightly overestimated for the HSE functional. However, as we have reported
previously, this is
not the case, {\em i.e.} the atomization energies of molecules
agree quite well with experiment,\cite{Paier} whereas those of solids
are underestimated.\cite{Paier2,PaierB3LYP} The reason for this is that the
inclusion of the exact exchange stabilizes spin polarized solutions
compared to non-spin polarized solutions, e.g.
spin polarized atoms are much more stable for hybrid functionals
than for semi-local exchange correlation functionals.
For $sp$ elements and $d$ metals this leads to a slight and sizeable underestimation of the
atomization energies, respectively.\cite{Paier2}
Neglecting these important spin-polarization effects,
hybrid functionals do behave reasonably similar to semi-local functionals,
with the tendency of increasing the cost for surfaces compared to the
corresponding semi-local functionals. 

Clearly, however, none of
the functionals can predict the surface energies and the
adsorption energies well at the same time.
From our point of view, the good prediction of the B3LYP and BLYP for adsorption energies
and the adsorption site is not related to a better description of the underlying physics
but rather accidental and fortuitous. By reducing the surface energy 
to an, in fact, unrealistically low value--- worsening the lattice constants---
the right site order is eventually recovered, but there is little doubt,
the reduction of the surface energy is not in accord with experiment.
We believe that the main error of semi-local functionals is in the description of
the strength of the back donation to the 2$\pi^*$ orbital which
tends to be overestimated using any of the semi-local density functionals.
Hybrid functionals improve on that aspect, but unfortunately simultaneously increase
the $d$ band width and concomitantly
the overall adsorption energies.

\section{Summary and Conclusions}
\label{sec:summary}

Overall, the situation is unsatisfactory. Considering CO adsorption
alone, the B3LYP and BLYP functionals seem to offer a very good description,
but the price to pay is that the description of the metals, in
particular $d$ metals, is unsatisfactory. Errors of 10 \% in the
volume   are unacceptable by today's standards, and worse, the
atomization energies of $d$ metals are wrong by up to 50 \% for  the
B3LYP functional.\cite{PaierB3LYP} Previously published seemingly
good B3LYP results\cite{Gil,Doll,Neef,Scheffler} must be
reconsidered, when the broader picture is taken into account. B3LYP and BLYP
{\em fail for metals and surface energies}: we feel there is hardly a point to  predict
surface properties, if the underlying substrate and surface is not accurately
and properly  described. 

The HSE functional performs
significantly better for metals, yielding a better overall
description of the equilibrium lattice constants than most
semi-local functionals. But again, atomization energies for
transition metals are wrong by typically 20 \%,\cite{Paier2} and the
overall description of CO on transition metal surfaces 
improves only little compared to the PBE functional. In particular, the
increased adsorption energies are unsatisfactory. Overall, we are
forced to conclude that  for metals and metal surfaces, hybrid
functionals are hardly a major step forward. In accord with our
previous conclusions,\cite{Stroppa07} we believe that the inclusion of a sizable amount of 
non-local exchange is not capturing the physics properly in metals.
The non-local exchange interaction is strongly screened in metals, both in
the long range as well as medium range. In combination
with a semi-local correlation functional, the hybrid HSE06 functional 
overestimates the exchange interactions at medium distances.\cite{Paier2,Stroppa07}

Finally, we have shown that trends from one to the other semi-local 
functional can be easily understood considering the cost for
creating a surface. For the recently suggested AM05 and PBEsol functionals, surfaces are, in agreement 
with experiment, fairly expensive favouring bulk-like behaviour, 
whereas rPBE and BLYP underestimate surface energies and shift the stability towards smaller
fragments (molecules and atoms).
It is really puzzling that AM05 and PBEsol yield good bulk lattice constants and surface energies
but overestimate adsorption energies, whereas 
rPBE and BLYP underestimate binding in solids and surface energies 
but describe adsorption energies well. A functional that could resolve both issues would be a tremendous step
forward but we firmly believe that semi-local functionals can not achieve this goal.

\section{Acknowledgments}
This work was supported by the Austrian {\em Fonds zur F\"orderung
 der wissenschaftlichen Forschung}. A.S. thanks J. Paier for interesting and fruitful discussions. Figure~\ref{fig:hollow}
has been done using the XCrySDen package.\cite{Xcry}

\section{Appendix. 
 Effect of pseudopotential approximation in
hybrid-functional calculations}

\begin{figure}
\epsfig{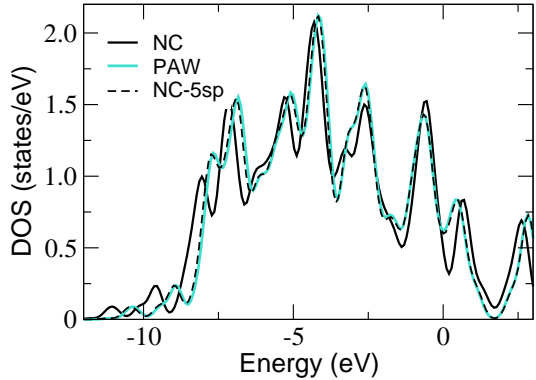}
\caption{ 
Electronic density of states (DOS) of Pt calculated using a
normconserving (NC) potential, a normconserving potential treating
the $5s$  and $5p$ as valence (NC$-5sp$) and using the PAW method for the PBEh
functional. The Brillouin zone was sampled using  $8\times8\times8$
k-points and a smearing width of $\sigma=0.3$\ eV was applied.}
\label{fig:1}
\end{figure}

In a recent work\cite{Trieste} it was suggested  that the
proper site preference  for CO on Pt(111) can be
obtained using the PBEh hybrid functional.
This observation contradicts our own finding,\cite{Stroppa07} and
we show here that the previous result is an artefact related to the pseudopotential
approximation applied in Ref.~\onlinecite{Trieste}.
To understand the discrepancy between our own work and that published in
 Ref.~\onlinecite{Trieste}, we generated a norm conserving  pseudopotential (NC-PP) mimicking
the FHI NC-PP applied in Ref.~\onlinecite{Trieste}.
Figure~\ref{fig:1} shows the density of states (DOS) evaluated using this normconserving
pseudopotential and the PAW method. In our implementation, the PAW method has
several advantages over NC-PPs. First, the exact
shape of the all-electron orbitals is restored avoiding any shape approximations in
the charge density, potentials and non-local exchange interaction.
Second, although the PAW method applied here is
using frozen core orbitals imported from a GGA calculation for the atom,
the exchange and correlation energy is re-evaluated  exactly and
consistently using all electrons (core and valence),
whereas in the pseudopotential approximation the hybrid functional is applied to
the valence electrons only, and the core-valence interaction is implicitly calculated
on the GGA level.
Clearly, both approaches result in a different density of states
for bulk Pt using the PBEh functional (see figure~\ref{fig:1}).
We note that no difference is observed on the GGA level, for which
the potentials were initially generated. Furthermore, we observe
that the PAW results are entirely robust and do not change upon
treating the $5s$ or $5p$ shell as valence (not shown), whereas
placing the $5s$ and $5p$ shell in the valence for the NC-PP (NC$-5sp$ in
figure~\ref{fig:1}) yields results that are practically identical to  the PAW
results.  The implicit treatment of the core-valence interaction on
the PBE level for the standard NC-PP causes fairly large errors.
\begin{table}
 \caption{
Adsorption energy of CO on Pt(111) evaluated for the top and fcc
site using a normconserving (NC) potential, a normconserving
potential treating the $5s$ and $5p$ as valence (NC$-5sp$) and using the PAW
method. The energy cutoff has been fixed to 1000, 1100 and 600 eV
respectively. Numbers in round brackets are taken from
Ref.~\onlinecite{Trieste} \label{tab:1}. }
\begin{ruledtabular}
\begin{tabular}{lccc}
          &    NC                       &   NC$-5sp$                  &    PAW              \\
PBEh top &    $-$2.07 ($-$1.80)         &     $-$1.92              &    $-$1.88        \\
PBEh fcc &    $-$1.85 ($-$1.67)         &     $-$1.89               &    $-$1.88         \\
\hline
PBE  top &  $-$1.56    ($-$1.53)         &  $-$1.56                  &   $-$1.54           \\
PBE  fcc &  $-$1.70    ($-$1.64)         &  $-$1.67                  &   $-$1.66           \\
\end{tabular}
\end{ruledtabular}
\end{table}


 In the second step we evaluated the CO adsorption energy
at the fcc and top site for a $\sqrt{3} \times \sqrt{3}$ slab using these
three potentials and   $4\times 4$  k-points   (similar technical
parameters were applied in Ref.~\onlinecite{Trieste}). The results are
summarized in table~\ref{tab:1}. At the PBE level, all three
potentials give very similar results to  those of
Ref.~\onlinecite{Trieste}. Large differences are observed at the PBEh
level. The standard NC potential yields the same site order as in
Ref.~\onlinecite{Trieste} (but different adsorption energies), whereas the NC$-5sp$ potential--- 
treating the
$5s$ and $5p$ electrons explicitly as valence ---and the PAW method show that the top and
fcc site are practically degenerate for this setup.
For a lower coverage using a $c(2\times4)$ surface cell and using more $k$ points,
the fcc site is consistently  preferred by 0.04 eV for the PBEh functional.
We conclude: the PBEh functional does not predict the right site
order on Pt(111) in the low coverage limit. 
Also the adsorption energy is significantly overestimated by the
hybrid PBEh functional, and \emph{overall the description is thus not
improved over the GGA case}.\cite{Stroppa07}
Furthermore, normconserving transition metal potentials generated using 
LDA/GGA functionals must be used with great care  in hybrid
functional calculations. We believe that this rule also applies to
elements beyond the third row ($4p$, $5p$ and $6p$
elements), where the core valence overlap
is often appreciable.


\begin{thebibliography}{}
\bibitem{LTRUNC} In order to accurately compute the $l-$decomposed charge, the projection of the wavefunctions onto
spherical waves has been performed setting the {\tt LTRUNC} parameter to 50 in the VASP code
\bibitem{bulkphase} We recall that   width, shape, and centroid position of the bulk $d$ bands
change at the surfaces mantaining the $d$ filling approximately constant
\end{thebibliography}
\end{document}